\documentclass[%
 reprint,
 superscriptaddress,
 amsmath,amssymb,
 aps,
 prx
floatfix,
]{revtex4-2}

\usepackage{graphicx}
\usepackage{dcolumn}
\usepackage{bm}
\usepackage{hyperref}
\usepackage{physics}

\begin{document}

\title{On-chip polarization-encoded single-qubit gates with twisted waveguides}
 
\author{Fyodor Morozko}
  \email{fedarm@post.bgu.ac.il}
  \affiliation{School of Electrical and Computer Engineering, Ben-Gurion University of the Negev, Beer-Sheva, Israel}
\author{Andrey Novitsky}
  \affiliation{Belarusian State University, Minsk, Belarus}
\author{Alexander Mikhalychev}
  \affiliation{B.~I.~Stepanov Institute of Physics, NAS of Belarus, Minsk, Belarus}
\author{Alina Karabchevsky}
  \affiliation{School of Electrical and Computer Engineering, Ben-Gurion University of the Negev, Beer-Sheva, Israel}

\begin{abstract}
  Integrated photonics is a remarkable platform for scalable classical and quantum light-based information processing.
  However, polarization manipulation on a chip despite of its fundamental significance in information processing remains elusive.
  Polarization manipulation capabilities have been recently demonstrated in femtosecond laser-inscribed twisted waveguides, although the systematic theoretical description of polarization manipulation has not been established for this architecture.
  In this work we develop a rigorous theory of a twisted waveguide unveiling its eigenmodes and transmission matrix in the closed form.
  Utilizing the developed theory, we demonstrate that twisted waveguides can realize virtually arbitrary polarization transformations while satisfying reasonable design constraints.
  This fact combined with low cost and ease of prototyping of laser inscribed photonic integrated circuits allows us to suggest twisted waveguide as a robust building block for on-chip polarization-encoded information processing.
\end{abstract}

\maketitle

\section{Introduction}
The seminal work by Knill, Laflamme, and Milburn (KLM)~\cite{ref:knill2001a}, where the authors proposed a scalable quantum computation protocol using purely linear optics, has boosted the exploration of photonics as a platform for implementing quantum information processing.
Since the first demonstration of the quantum controlled-NOT (CNOT) gate using the KLM protocol on silicon-on-silica chip in 2008 by Politi~\emph{et al}~\cite{ref:politi2008}
integrated photonics is considered on of the most promising platforms for implementing scalable quantum information processing due to its flexibility in light manipulation in a highly controllable manner~\cite{ref:slussarenko2019,ref:wang2020} and has already reached the level of maturity to allow creating large-scale reconfigurable quantum circuits involving a dozen of qubits~\cite{ref:carolan2015}.

To encode information in a single photon one must use its physical degrees of freedom such as path, momentum, angular momentum, and polarization.
For reaching a higher information processing capability per chip footprint it is desirable to make use of the maximum possible number of them.
Photon polarization is an always-available natural degree of freedom and is thus among the most widely used encoding mechanisms.
To benefit from using an integrated platform it is crucial to perform most or ideally all light manipulations on a chip as most losses occur at a stage of coupling light into a chip or from a chip.
However, despite the recognized strength of integrated photonics in controlling light, manipulation of polarization on a chip yet remains elusive.
Although integrated photonic polarization-encoded CNOT gate has been demonstrated in laser-written chips, the polarization manipulation in the reported works~\cite{ref:sansoni2010,ref:crespi2011,ref:zeuner2018} was performed using either bulk or fiber optics.
On-chip polarization manipulation schemes based on tilted basis waveguides are typically used serving as waveplates~\cite{ref:corrielli2014,ref:heilmann2015} where the waveguide symmetry axis exhibits the optical axis.
Such schemes, however, suffer from a number of drawbacks: they are extremely sensitive to fabrication tolerances and tend to have significant coupling losses due to the cross-section mismatch with normal waveguides ~\cite{ref:baier2018}.

It was already recognized in 1979 by~Ulrich~and~Simon that nontrivial polarization dynamics occur in twisted birefringent fibers~\cite{ref:ulrich1979} due to the interplay of linear and circular birefringence, where the former is caused, e.g., by core ellipticity or stress, while the latter is the topological effect owing to twisting as was discussed also in the later works~\cite{ref:michie2007,ref:argyros2009}.
Due to the recent advances in integrated photonics fabrication technology, especially, in the laser writing, the integrated photonic twisted waveguides have become a reality and have already been suggested as broadband adiabatic polarization rotators~\cite{ref:schumann2014,ref:hou2019,ref:sun2022a}.
In the above mentioned works several effects were observed in twisted waveguides which cannot be easily explained by adiabatic mode evolution principle: namely, decrease of polarization conversion efficiency with increase of the twist length in~\cite{ref:hou2019} and spectral oscillations of polarization conversion efficiency in~\cite{ref:sun2022a}.
The explanation of these effects calls for a rigorous theoretical description of polarization dynamics in a twisted waveguide which we develop in present work.
As our analysis shows, twisted waveguides are capable not only of linear polarization rotations but of arbitrary polarization transformations aiming to optical activity arising from the interplay of structural linear and circular birefringence.
This allows to suggest twisted waveguides as on-chip elliptical waveplates capable of implementing arbitrary unitary operations in polarization-encoded quantum and classical information processing circuits.
The analytical model we develop here may significantly facilitate the prototyping of twisted waveguide-based devices since it provides a deep insight on the underlying physics and allows to quickly perform multi-parameter optimizations.
Furthermore, being fabricated with the femtosecond laser inscription technology, which combines low-cost fabless and maskless fabrication process with excellent design flexibility~\cite{ref:marshall2009,ref:sansoni2010,ref:crespi2011,ref:meany2015,ref:flamini2015,ref:heilmann2015,ref:zeuner2018,ref:sun2022a}, twisted waveguide exhibits a promising building block for experimental realization of polarization-encoded quantum information processing integrated circuits.

\section{Coupled mode theory for twisted waveguide}
\label{sec:cmt}

\begin{figure}[htpb]
  \centering
  \includegraphics[width=\linewidth]{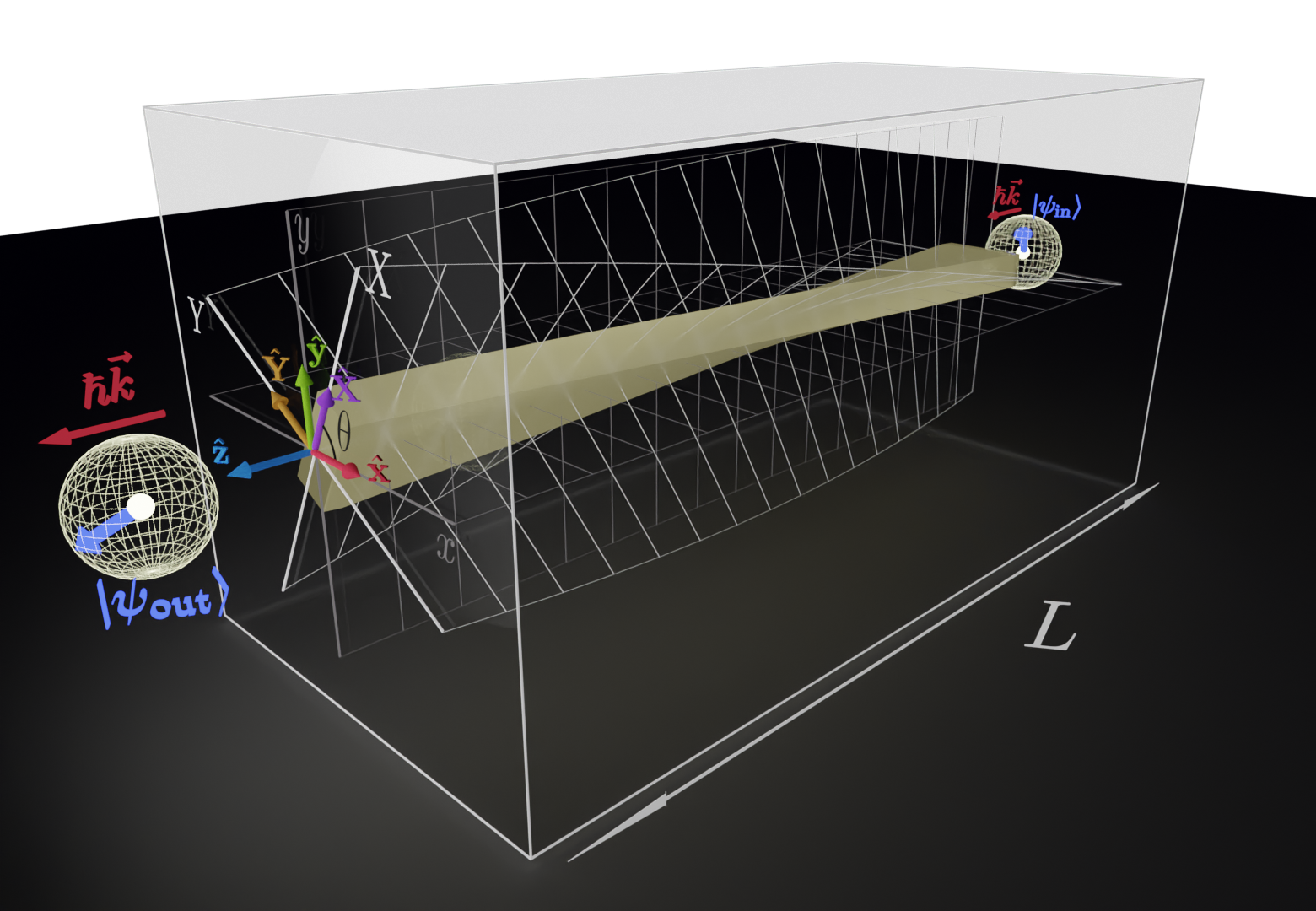}
  \caption{Schematic of a twisted waveguide realizing polarization-encoded single-qubit gate.
    $L$ is the twist length, $\theta$ is the twist angle.
    Input and output single-photon states $\ket{\psi_\mathrm{in}}$ and $\ket{\psi_\mathrm{out}}$ are represented as positions on the Bloch sphere.
}
  \label{fig:scheme}
\end{figure}
Expressing Maxwell's equations in a helical reference frame and taking into account the helical symmetry of a twisted waveguide one can obtain an operator equation~\cite{ref:weiss2013}
\begin{equation}
  i\pdv{z}\vec F=G\vec F,
  \label{eq:evolution}
\end{equation}
where $\vec F$ is a four-component vector of transverse electric and magnetic fields, $G$ is a $z$-independent evolution operator.
It is possible to separate variables in Eq.~\eqref{eq:evolution} as $\vec F(\vb{r})=\vec f(\vb{r}_\perp)e^{-i\beta z}$, where the radius-vector $\vb{r}$ is expanded in transversal and longitudinal components as $\vb{r}=\vb{r}_\perp+z\vu{z}$ with the transverse radius-vector $\vb{r}_\perp$ expanded in helical basis as $\vb{r}_\perp=X\vu{X}+Y\vu{Y}$.
The transverse basis vectors $\{\vu{X},\vu{Y}\}$ of the helical reference frame as well as the basis vectors $\{\vu{x},\vu{y},\vu{z}\}$ of the Cartesian laboratory frame are depicted in Fig.~\ref{fig:scheme}.
The separation of variables leads to the eigenmode equation for the vector $\vec f(\vb{r}_\perp)$ and the eigenvalue $\beta$.
For the waveguide composed of isotropic materials one can derive the eigenmode equation in an alternative way: from the wave equation in terms of either magnetic or electric fields eliminate the longitudinal field obtaining the eigenmode equation for the transverse fields.
Such formulated eigenmode equation has dimensionality of two giving benefits for numerical implementation.
We write the eigenmode equation for the transverse electric field as
\begin{equation}
  L(\beta,\alpha)\ket{\psi}=0,
  \label{eq:twisted-eigenproblem-general}
\end{equation}
where $L(\beta,\alpha)$ is the equation operator quadratic both in the eigenvalue $\beta$, and the twist rate $\alpha=\theta/L$, where $\theta$ and $L$ are the twist angle and length, respectively, as shown in Fig.~\ref{fig:scheme}.
We denote by $\ket{\psi}$ the transverse modal electric field expanded in helical frame as $\braket{r_\perp}{\psi}=e_X\vu{X}+e_Y\vu{Y}$ while using ket- and bra-vectors as a convenient notation for classical modes.
The classical modes, however, have tight relation to the quantum states as we point out in the following section.
The operator $L(\beta,\alpha)$ reads as
\begin{equation}
  L(\beta,\alpha)=A-\beta^2+\alpha V(\beta)+\alpha^2D\quad 
  \label{eq:operatorL}
\end{equation}
with
\begin{equation}
  V(\beta)=2i(\beta^{-1}B+\beta C),
  \label{eq:operatorV}
\end{equation}
where $A$, $B$, $C$, $D$ are the two-dimensional operators depending on $X$, $Y$ and their derivatives with $A=H_0$ coinciding with the eigenvalue equation operator for an untwisted waveguide in the laboratory frame~\cite{ref:morozko2022}.
See Appendix~\ref{sec:derivation} and the reference cited for the derivation of Eqs.~(\ref{eq:twisted-eigenproblem-general}-\ref{eq:operatorV}) from Maxwell's equations in covariant formulation.

The operator $L$ in Eq.~\eqref{eq:operatorL} poses a polynomial eigenvalue problem which, despite that it can be solved directly~\cite{ref:mehrmann2004}, we solve using perturbative approach with respect to the small twist rate $\alpha$.
The perturbative approach allows to convert the polynomial eigenmode equation to the standard, i.e., linear eigenvalue equation and to express the eigenmodes explicitly as functions of eigenmodes of an unperturbed (untwisted) waveguide and the twist rate $\alpha$.
Similar perturbative solutions were obtained earlier for twisted microwave waveguides~\cite{ref:yabe1984} and helical fibers~\cite{ref:ma2011b,ref:weiss2013}.
However, those formulations relied on the modal expansions peculiar to the studied waveguide cross-sectional configuration and are not directly applicable for the twisted waveguides of arbitrary profile.

In this section, we outline a perturbative theory of twisted waveguides which makes no assumptions on the waveguide geometry and relies solely on the fundamental property of orthogonality of the guided modes.
Considering the twist rate as a small perturbation parameter we can establish two simplifications.
First, we approximate $V(\beta)$ in Eq.~\eqref{eq:operatorL} with $V(\beta_0)$, where $\beta_0$ is the unperturbed eigenvalue. Second, we omit the quadratic term $\alpha^2D$.
With these simplifications in hand we formulate a linearized perturbative eigenvalue problem as follows
\begin{equation}
  H\ket{\psi}=\beta^2\ket{\psi},\quad H=H_0+\alpha V(\beta_0).
  \label{eq:twisted-eigenproblem-pert}
\end{equation}
Since the eigenvalues of guided modes are very close, especially for polarization modes, the normal perturbation theory fails~\cite{ref:kato1995}.
It is however possible to find the perturbed modes in terms of the coupled mode theory looking for them in the form of linear combinations of the unperturbed modes 
\begin{equation}
  \ket{\tau_\nu}=\sum_{\mu=1}^{N}M_{\mu\nu}\ket{\mu},
  \label{eq:tau-M-hv}
\end{equation}
where columns of $M_{\mu\nu}$ correspond to expansion coefficients of the $\nu$-th twisted mode $\ket{\psi}=\ket{\tau_\nu}$ over untwisted modes $\ket{\mu}$ with $N$ being the number of modes.

By substituting Eq.~\eqref{eq:tau-M-hv} into Eq.~\eqref{eq:twisted-eigenproblem-pert} and using orthogonality of the unperturbed modes $\braket{\mu}{\nu}=\delta_{\mu\nu}$, we reduce the problem of finding perturbed modes $\ket{\tau_\nu}$ to the problem of diagonalization of the Hamiltonian matrix $H_{\mu\nu}=\matrixel{\mu}{H}{\nu}$ in the basis of the unperturbed modes, while the matrix elements constitute overlap-type integrals, as we show in Appendix~\ref{sec:matrixel}.

Perturbation operator $V$ and inner products constituting matrix elements can be calculated analytically if the unperturbed modes are either known analytically or possess symmetries.
In a general case the perturbation operator can be implemented numerically by approximating the derivatives with matrices using Finite Difference Method while the unperturbed modes can be also calculated with Finite Difference Method or some other method such as Finite Element Method.

\section{Single-mode twisted waveguide}
\begin{figure}[htpb] 
  \includegraphics[width=\linewidth]{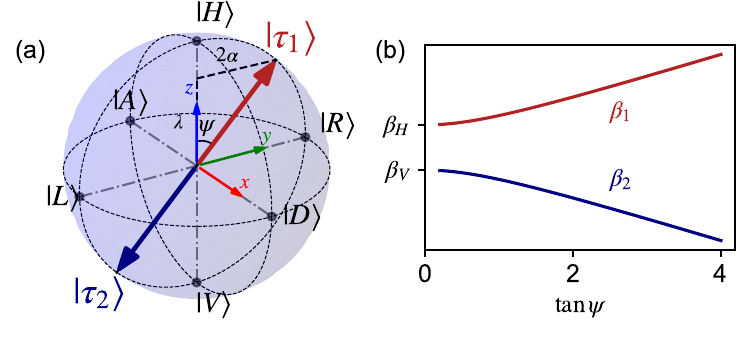}
  \caption{
  Eigenvectors (a) and eigenvalues (b) of the eigenmodes of a twisted waveguide as a function of twist rate expressed in terms of the angle $\psi$ defined in Eq.~\eqref{eq:psi}} 
  \label{fig:eigenmodes}
\end{figure}

It is instructive to consider a single-mode twisted waveguide since in this case, it is possible to find matrix elements $\matrixel{\mu}{V}{\nu}$ explicitly imposing the following reasonable assumptions.
Firstly, if the waveguide's cross-section defined by permittivity profile $\varepsilon(X,Y)$ is rectangular or, more generally, the function $\varepsilon(X,Y)$ is even with respect to both $X$ and $Y$, the modal profiles must be either even or odd functions of $X$ and $Y$~\cite{ref:xiong2017}.
Secondly, we consider that twist does not cause mode leakage, so that the matrix $V$ as well as the Hamiltonian matrix $H$ are Hermitian.
Finally, we assume that the twisting axis coincides with the center of the waveguide, that is, its symmetry in helical coordinates is unbroken.
Then the symmetries restrict integrands in diagonal matrix elements $\matrixel{\mu}{V}{\mu}$ to be odd functions causing these matrix elements to vanish while the off-diagonal elements are complex conjugates to each other $\matrixel{1}{V}{2}=\matrixel{2}{V}{1}^*$.
By calculating the only remaining nontrivial matrix element $\matrixel{1}{V}{2}$ we reveal that the matrix $V$ is proportional to the Pauli-$Y$ matrix
\begin{equation}
  \matrixel{\mu}{V}{\nu}=
  \begin{pmatrix}
    0&-2i\bar\beta\\
    2i\bar\beta&0 
  \end{pmatrix}
  =2\bar\beta\sigma_y,
  \label{eq:V-HV}
\end{equation}
where $\bar\beta=(\beta_H+\beta_V)/2$ is the average propagation constant.
The Hamiltonian matrix 
\begin{equation}
  \matrixel{\mu}{H}{\nu}=
  \begin{pmatrix} 
    \beta_H^2&-2i\alpha\bar\beta\\
    2i\alpha\bar\beta&\beta_V^2
  \end{pmatrix}
\end{equation}
has a couple of eigenvalues $\beta_{1,2}=\bar\beta\pm\frac{1}{2}\sqrt{\lambda^2+4\alpha^2}$.
Here $\lambda=\beta_H-\beta_V$ is the linear birefringence in the untwisted waveguide reciprocal to the linear beat length as $\lambda=2\pi/L_B$.
Introducing an angle $\psi$ according to the definition
\begin{equation}
  \tan\psi=2\alpha/\lambda,
  \label{eq:psi}
\end{equation}
one can represent eigenvalues and eigenvectors in the compact form as
\begin{equation}
  \beta_{1,2}=\bar\beta\pm\frac{\lambda}{2\cos\psi}.
  \label{eq:beta-12}
\end{equation}
and
\begin{align}
  \ket{\tau_1}&=\cos\psi/2\ket{H}+i\sin\psi/2\ket{V},\\
  \ket{\tau_2}&=i\sin\psi/2\ket{H}+\cos\psi/2\ket{V},
  \label{eq:tau-12}
\end{align}
respectively. 
If a single-photon state is supplied to a twisted waveguide, the vectors $\ket{\tau_{1,2}}$ can be naturally associated with polarization qubits.
It is convenient to illustrate these states on the Bloch sphere as shown in Fig.~\ref{fig:eigenmodes}(a).
Propagation constants \eqref{eq:beta-12} are shown in Fig.~\ref{fig:eigenmodes}(b). 
By comparing the expansions above with~\eqref{eq:tau-M-hv} it is easy to see that 
\begin{equation}
  M=\begin{pmatrix} 
    \cos\psi/2&i\sin\psi/2\\
    i\sin\psi/2&\cos\psi/2\\
  \end{pmatrix}=\exp(i\sigma_x\psi/2)
  \label{eq:M}
\end{equation}
is the rotation matrix corresponding to rotation of the Bloch sphere around the $x$ axis by the angle $\psi$, where $\sigma_x$ is the Pauli-$X$ matrix. 
The expression (\ref{eq:M}) of the matrix $M$ in terms of the Pauli matrix reveals the geometric meaning of $\psi$ visualized in Fig.~\ref{fig:eigenmodes}(a).
It should be noticed that rotation of the eigenmodes around the $x$ axis is caused by interplay of the linear birefringence $\lambda$ induced by unequal cross-section dimensions and topological circular birefringence $2\alpha$ induced by twisting~\cite{ref:ulrich1979,ref:michie2007}.

At very slow twist rates, namely, when $\alpha\ll\lambda$, the angle $\psi$ is close to $0$ and the twisted waveguide modes coincide with horizontally $\ket{H}$ and vertically $\ket{V}$ polarized modes of the untwisted waveguide.
On the other hand, at the rapid twist rates, when $\alpha\gg\lambda$, $\psi$ approaches $\pi/2$ and the modes become circularly polarized as~$\ket{R}$~and~$\ket{L}$ in Fig.~\ref{fig:eigenmodes}(a).
In any intermediate case, the polarization is elliptical being the mixture of linear and circular contributions.
Noteworthy, in the case of zero linear birefringence our theory reproduces linear eigenvalue separation $\beta_1-\beta_2=2\alpha$ as predicted by degenerate perturbation theory isomorphic to the theory of Zeeman effect in the weak magnetic field~\cite{ref:weiss2013}.

\section{Transmission matrix of a single-mode twisted waveguide}
Knowing that a single-mode twisted waveguide operates as an elliptical waveplate we can proceed to the derivation of its Jones matrix or transmission matrix $T$ in the waveguide terminology
\begin{equation}
  T: \ket{\psi_\mathrm{out}}=T\ket{\psi_\mathrm{in}},
\end{equation}
where $\ket{\psi_\mathrm{in(out)}}=H_\mathrm{in(out)}\ket{H}+V_\mathrm{in(out)}\ket{V}$ are the input (output) polarization states as shown in Fig.~\ref{fig:scheme}.
We immediately observe that since the twisted modes $\ket{\tau_{1, 2}}$ are the eigenmodes of $T$, the transmission matrix in their basis is diagonal $\matrixel{\tau_\nu}{T}{\tau_\mu}=\exp(-i\beta_{\mu}L)\delta_{\mu\nu}$.
After extracting the unimportant global phase factor $\exp(-i\bar\beta L)$ we find that $\matrixel{\tau_\nu}{T}{\tau_\mu}=\exp(-i\sigma_z\phi/2)$, where $\phi=\delta\beta L$ ($\delta\beta=\beta_1-\beta_2$) is the accumulated phase difference between the modes or retardance.
Using Eq. \eqref{eq:beta-12} the retardance can be written by means of angle $\psi$ and linear birefringence as $\phi=\lambda L/\cos{\psi}$.
Positions of the vectors $\ket{\tau_\mu}$ on the Bloch sphere are unchanged upon transformation $T$. This means that they lie on the axis of the rotation as depicted in Fig.~\ref{fig:eigenmodes}(a).
These geometrical considerations allow us to immediately find the matrix elements of $T$ in the basis $\{\ket{H},\ket{V}\}$;
\begin{equation}
  \matrixel{\nu}{T}{\mu}=\exp(-i\vu{m}\cdot\vec{\sigma}\phi/2),
  \label{eq:T-hv}
\end{equation}
where $\vu{m}=\vu{z}\cos{\psi}+\vu{y}\sin{\psi}$ is the unit vector along $\ket{\tau_1}$, $\vec\sigma=\{\sigma_x,\sigma_y,\sigma_z\}$ is the vector of Pauli matrices, 
 This result can be also obtained algebraically using the unitary transformation $\matrixel{\tau_\nu}{T}{\tau_\mu}=\matrixel{\nu}{{M}^{\dagger}TM}{\mu}$.
 We provide algebraic derivation of~Eq.~\eqref{eq:T-hv} in Appendix~\ref{sec:algebr-T}.
We emphasize that matrix elements \eqref{eq:T-hv} defined so far  refer to the helical reference frame.
To obtain transmission matrix in the laboratory frame we relate the components in different bases $\ket{\mu}=J\ket{\mu'}$ by means of the Jacobian matrix  $J=\exp(-i\sigma_y\alpha z)$, so that
\begin{equation}
  \matrixel{\nu'}{T}{\mu'}=\exp(i\sigma_y\theta)\exp(-i\vu{m}\cdot\vec{\sigma}\phi/2)
\end{equation}
are the matrix elements in the laboratory frame.

Therefore, transformation of polarization can be treated as a composition of two subsequent Bloch sphere rotations: first, rotation about the $\ket{\tau_\mu}$ line as seen from the helical reference frame and, second, rotation of the basis vectors of the helical frame about the $y$ axis of the Bloch sphere by $2\theta$.
To express $T$ as a single rotation we compose two rotations $\exp(i\sigma_y\theta)$ and $\exp(-i\vu{m}\cdot\vec\sigma\phi/2)$ as described in~\cite{ref:nielsen2010}
\begin{equation}
  T=\exp(-i\vu{n}\cdot\vec{\sigma}\chi/2),
  \label{eq:Trot}
\end{equation}
where the angle $\chi$ and the axis $\vu{n}$ are defined by the waveguide parameters $\theta$, $\psi$, and $\phi$ as
\begin{align}
  \cos\frac{\chi}{2}&=\cos\theta\cos\frac{\phi}{2}+\sin\theta\sin\frac{\phi}{2}\sin\psi,\nonumber\\
  n_x\sin\frac{\chi}{2}&=\cos{\psi}\sin{\theta}\sin\frac{\phi}{2},\nonumber\\
  n_y\sin\frac{\chi}{2}&=\cos{\frac{\phi}{2}}\sin{\theta}
  +\cos{\theta}\sin\frac{\phi}{2}\sin{\psi},
  \label{eq:angles}
  \\
  n_z\sin\frac{\chi}{2}&=\cos{\theta}\cos{\psi}\sin{\frac{\phi}{2}}.\nonumber
\end{align}

In the slow twisting regime ($\psi\to0$), the gate~$T$, as follows from~Eq.\eqref{eq:angles}, reduces to the rotation about the $y$ axis by the angle $2\theta$: $T\to\exp(-i\sigma_y\theta)$.
Interestingly, in Ref.~\cite{ref:hou2019} the authors observed slight ripples in polarization conversion dependence on the twist length, the amplitude being larger at smaller twist lengths.
Those ripples can be explained by departure from the linear birefringence regime equivalent to the presence of effective optical activity ($\psi\neq0$).
Such an effect can be also related to the spectral oscillations of polarization conversion efficiency experimentally observed for laser-inscribed twisted waveguides in Ref.~\cite{ref:sun2022a}: the spectral dependency of polarization conversion arises due to significant circular birefringence and inherits an elliptical waveplate effect in low birefringence borosilicate glass platform.
In the case of dominant circular birefringence ($\psi\to\pi/2$) we can see that the gate reduces to unity, $T\to1$, because the optical activity in helical fibers appears only if the fiber possesses some linear birefringence.
That is why a square twisted waveguide would not affect the polarization.
The absence of polarization conversion may have a useful application: one can insert short mode adapters at the facet of the twisted waveguide to compensate for cross-section mismatch (which may appear if one uses oblique twist angles) without affecting the gate performance.

\section{Arbitrary waveplates and single-qubit gates}
In this section, we are going to estimate the capability of twisted waveguides to perform arbitrary polarization transformations.
To estimate this capability we posed an inverse design problem: For a given target unitary operator,
namely, a Bloch sphere rotation with given Euler angles find the twisted waveguide realization defined by two parameters: the length $L$ in terms of linear beat lengths $L_B$ and the twist angle $\theta$.
To quantify the quality of realized operations, we use single-qubit gate fidelity measure $F=\frac{1}{2}+\frac{1}{12}\sum_{j=x,y,z}\Tr(T\sigma_j{T}^{\dagger}U\sigma_j{U}^{\dagger})$ proposed in Ref.~\cite{ref:bowdrey2002}, where $U$ is the target gate and $T$ is its twisted waveguide realization as defined~by~\eqref{eq:Trot}~and~\eqref{eq:angles}.
$F$ measures the average deviation of the $\ket{\psi_\mathrm{out}}^\mathrm{actual}=T\ket{\psi_\mathrm{in}}$ from the target state $\ket{\psi_\mathrm{out}^\mathrm{target}}=\ket{\psi_\mathrm{in}}$ among a set of all possible input states $\ket{\psi_\mathrm{in}}$.
We first parametrize rotation axis $\vu{n}$ in Eq.~\eqref{eq:angles} in spherical coordinates $\vartheta$ and $\varphi$ as $\vu{n}=\sin\vartheta\cos\varphi\vu{x}+\sin{\vartheta}\sin{\varphi}\vu{y}+\cos{\vartheta}\vu{z}$ establishing the relation between the Euler angles $\{\vartheta,\varphi,\chi\}$ and design parameters $\{\theta,\psi,\phi\}$
\begin{align}
  \cos\frac{\chi}{2}&=\cos\theta\cos\frac{\phi}{2}+\sin\theta\sin\frac{\phi}{2}\sin\psi,\nonumber\\
  \sin{\vartheta}\cos{\varphi}\sin\frac{\chi}{2}&=\cos{\psi}\sin{\theta}\sin\frac{\phi}{2},\nonumber\\
  \sin{\vartheta}\cos{\varphi}\sin\frac{\chi}{2}&=\cos{\frac{\phi}{2}}\sin{\theta}
  +\cos{\theta}\sin\frac{\phi}{2}\sin{\psi},
  \\
  \cos{\vartheta}\sin\frac{\chi}{2}&=\cos{\theta}\cos{\psi}\sin{\frac{\phi}{2}}.\nonumber
\end{align}
Then we sweep $\varphi$ and $\chi$ from $0$ to $2\pi$ and $\vartheta$ from $0$ to $\pi$ effectively covering a discrete grid of all possible single qubit gates.
The grid dimensions were $33\times65\times17$ for respectively polar, azimuthal, and rotation angles, thus, giving the total number of target gates equal $36465$.
In order the optimization results to be of practical significance we constrained the maximum twisted waveguide length $L$ and twist angle $\theta$ and calculated the worst fidelity $F_{\mathrm{min}}$ over a set of target operators for different values of the constraints $L_\mathrm{max}$, $\theta_\mathrm{max}$.
We provide details of the numerical solution to the inverse design problem in Appendix~\ref{sec:inverse-design}.
Fig.~\ref{fig:Fworst} summarizes the results of our analysis.
Fig.~\ref{fig:Fworst}(a) shows the worst fidelity $F_\mathrm{min}$ among all the designs.
One can see from Fig.~\ref{fig:Fworst} monotonic improvement of $F_\mathrm{min}$ with respect to increase of both $\theta_\mathrm{max}$ and $L_\mathrm{max}$ guarantying, for instance, fidelity greater than $0.95$ for $L_\mathrm{max}>5L_B$ and $\theta_\mathrm{max}>5\pi$ for any operation.
The worst fidelities for the approximations of rotations around different axes for~$\theta_\mathrm{max}=20\pi$ and three values of $L$ are illustrated in Figs.~\ref{fig:Fworst}(b-d).
Position of points on spheres in these figures is associated with the rotation axis while the colors correspond to the worst fidelity across a set of angles of rotation [$\chi$~in~Eq.\eqref{eq:Trot}] with respect to this axis.
Bar charts in Figs.~\ref{fig:Fworst}(e-g) show the fidelity distributions.
For the considered constraints the fidelities appear to group near the unity, whereas increasing the maximum length $L$ narrows the distribution.
The results demonstrate that the absolute majority of gates can be approximated with fidelity $>0.95$ while the twisted waveguides are restricted to a few linear beat lengths with the twist angle of~$20\pi$ ($10$ total twists).
State-of-the-art laser-written waveguides typically exhibit birefringences $\delta n \sim 10^{-5}\mathrm{-}10^{-4}$ in terms of modal indices depending on the particular fabrication process and cross-section dimensions.
Such a birefringence ensures the linear beat length $L_B \sim 8-0.8~\mathrm{cm}$ at wavelength $800~\mathrm{nm}$~\cite{ref:sun2022a}.
We thus can conclude that laser-written twisted waveguides implementing any possible single qubit gate have comparable sizes to the laser-written architectures reported earlier~\cite{ref:crespi2011,ref:heilmann2015}.
\begin{figure*}[htpb]
  \centering
  \includegraphics[width=0.8\linewidth]{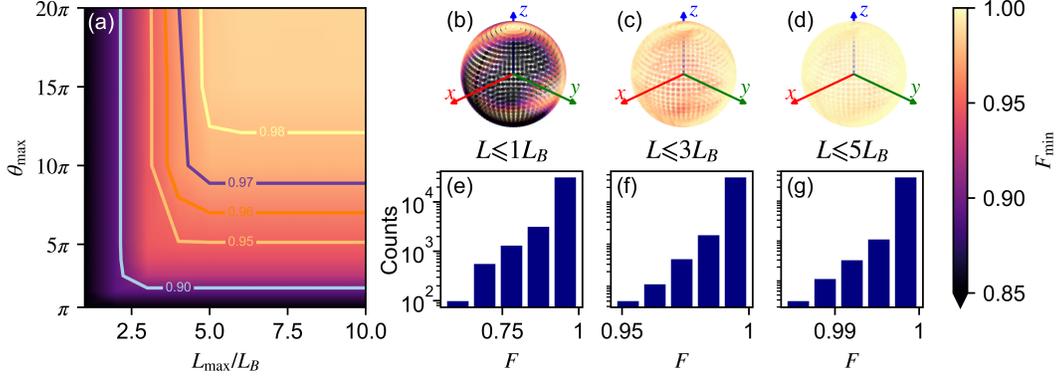}
  \caption{Twisted waveguide approximation of arbitrary single-qubit gates.
    (a) shows worst fidelity $F_\mathrm{min}$ overall single qubit gates as a function of twisted waveguide design constraints,
  where $\theta_\mathrm{max}$ is the maximum twist angle, $L_\mathrm{max}$ is the maximum twist length measured in terms of linear beat lengths $L_B$.
(b-d) show the worst fidelity over rotations around a given axis with $\theta_\mathrm{max}=20\pi$ and three different $L_\mathrm{max}$ constraints, histograms (e-g) below the spheres visualize the distribution of fidelities.}
  \label{fig:Fworst}
\end{figure*}

A possible way to reduce the sizes of gates is to stack several twisted waveguides.
In this case one has greater number of design parameters which allows to reach higher fidelities at more compact sizes.
For instance, a single twisted waveguide implementing Pauli-$X$ operation with fidelity $0.992$ has the length $4.473L_B$ while a waveguide composed of two stacked waveguides – a $\pi/2$-twisted waveguide of the length $0.866L_B$ implementing Pauli-$Y$ operation and a straight waveguide of the length $0.5L_B$ implementing Pauli-$Z$ operation – has the total length of $1.366L_B$ while implementing Pauli-$X$ with unit fidelity.

\section{Conclusion}
To sum up, we have developed a perturbative theory for a twisted waveguide which is applicable to twisted waveguides of arbitrary cross-section.
Within the perturbation theory we have presented the mode fields of a twisted waveguide as explicit functions of the mode fields of an untwisted waveguide and the twist rate. This result is of significant value both in numerical and analytical study of twisted waveguides.
In order to gain a clear insight to the physics in twisted waveguides we have applied the developed theory to a single-mode rectangular twisted waveguide. 
For such a system we have revealed analytical expressions for eigenmodes and transmission matrix in terms of the Pauli matrices. These compact equations vividly show the interplay of linear and circular birefringence and, thus, are useful for implementing elliptical waveplate operations.
Performing a parameter optimization of the transmission matrix we have demonstrated that a single-mode twisted waveguide can realize arbitrary polarization operations while obeying reasonable design constraints which readily allows implementation of arbitrary single-qubit maps on polarization qubits.
Although in this work we have analyzed only single-qubit operations, the potential of twisted waveguides in quantum information processing is not limited by them: it is possible to realize multi-qubit maps using composite structures such as coupled twisted waveguides.
Since the multi-qubit gates typically involve single-qubit operations as compensation schemes, the availability of arbitrary single qubit operations is extremely useful.
Furthermore, the developed theory allows expressing the modes and transmission matrices of coupled twisted waveguide structures in the framework of the standard coupled mode theory, that is, in terms of individual waveguide modes which significantly simplifies design of coupled twisted waveguide-based quantum gates.

Twisted waveguide, thereby, exhibits a promising on-chip polarization-manipulation building block.
We believe that our research may encourage optical community for adopting twisted waveguides as a novel paradigm to realize polarization-encoded quantum and classical information processing circuits, particularly with the low-cost laser inscription technology.

\section*{Acknowledgements}
A.K. acknowledges the support of the Israel Science foundation ISF Grant 
No. 2598/20 and the EU ERA-NET, Ministry of Energy, Grant No. 221-11-032

\appendix

\section{Derivation of the eigenmode equation}
\label{sec:derivation}
In this section we briefly overview the derivation of the eigenmode equation for a twisted waveguide given by Eqs.~(\ref{eq:twisted-eigenproblem-general}-\ref{eq:operatorV}).
See Ref.~\cite{ref:morozko2022} and references there for more details.

We start with the source-free Maxwell's equations in covariant form 
\begin{align}
  &\epsilon^{ijk}\nabla_{j}H_{k}=-ik_0\varepsilon E^{i}
  \label{eq:maxwell-rotH}
  \\
  &\epsilon^{ijk}\nabla_{j}E_{k}=ik_0H^{i},
  \label{eq:maxwell-rotE}
  \\
  &\nabla_i\varepsilon E^i=0,
  \label{eq:maxwell-divE}
  \\
  &\nabla_i H^i=0,
  \label{eq:maxwell-divH}
\end{align}
here, $\epsilon^{ijk}$ is the fully antisymmetric (Levi-Chivitta) tensor in three dimensions, $\nabla_i$ denotes the covariant derivative, $\varepsilon$ is the permittivity function, $k_0$ is the vacuum wavenumber.
From Eqs.~(\ref{eq:maxwell-rotH}-\ref{eq:maxwell-rotE}) one can derive the wave equation for the contravariant electric field
\begin{equation}
  g^{kl}\nabla_k\nabla_lE^i-g^{ik}\nabla_k(\nabla_lE^l)+k_0^2\varepsilon E^{i}=0,
  \label{eq:waveE}
\end{equation}
where $g^{kl}$ is the contravariant metric tensor.
Introducing modal ansatz $E^{i}=e^{i}\exp(-i\beta Z)$ into \eqref{eq:waveE} and
Eq.~(\ref{eq:maxwell-divE}) and using independence of the twisted-waveguide's permittivity function on the longitudinal coordinate in helical coordinates $\nabla_3\varepsilon=0$ we eliminate the longitudinal contravariant component $E^3$ from~\eqref{eq:waveE}.

Then, substituting the metric tensor in helical basis
\begin{equation}
  g^{ik}=\begin{pmatrix}
    1+\alpha^2Y^2&-\alpha^2XY&\alpha{Y}\\
    -\alpha^2XY&1+\alpha^2X^2&-\alpha{X}\\
    \alpha{Y}&-\alpha{X}&1
  \end{pmatrix}
\label{eq:contra_metric}
\end{equation}
and performing covariant differentiation, we establish the eigenproblem in the form given by Eqs.~(\ref{eq:twisted-eigenproblem-general}-\ref{eq:operatorV}) with operators $A$, $B$, $C$, $D$ defined in~\cite{ref:morozko2022}.

\section{Calculating matrix elements of perturbation operator}
\label{sec:matrixel}
To calculate matrix elements $\matrixel{\mu}{V}{\nu}$ we insert the identity operator $I=\int\ketbra{r_\perp}d^2r_\perp$ 
\begin{equation}
  \matrixel{\mu}{V}{\nu}=
  \matrixel{\mu}{IVI}{\nu}
  =\int_{A_\infty}\braket{\mu}{r_\perp'}\matrixel{r_{\perp}'}{V}{r_\perp}\braket{r_\perp}{\nu}d^2r_\perp'd^2r_\perp.
  \label{eq:V-overlap}
\end{equation}
The modes $\ket{\mu}$ obey orthogonality relation
\begin{equation}
  \braket{\mu}{\nu}=\int_{A_\infty}\epsilon^{3ij}h^{*}_{j,\mu}e_{i,\nu}d^2\vb{r}_\perp=\delta_{\mu\nu},
\end{equation}
so in order to obtain a correct inner product satisfying orthogonality the bra-vector $\bra{\mu}$ must be chosen as
\begin{equation}
  \braket{\mu}{r_\perp}=\epsilon^{3ij}h^{*}_{j,\mu}(\vb{r}_\perp)g_{ik},
\end{equation}
where asterisk $(\cdot)^{*}$ denotes complex conjugation.

As we discuss in the main text, in the case of single-mode twisted waveguide with even permittivity function $\varepsilon(X,Y)$ the symmetries of the modal functions cause matrix elements of the operators $B$ and $C$ in the modal basis $\{\ket{H},\ket{V}\}$ to reduce to
\begin{equation}
  \matrixel{\mu}{B}{\nu}=0,\;
  \matrixel{\mu}{C}{\nu}
  =\begin{pmatrix} 0&-1\\1&0 \end{pmatrix}
\end{equation}
inducing~Eq.~\eqref{eq:V-HV} for the matrix $\matrixel{\mu}{V}{\nu}$.

\section{Algebraic derivation of the single-mode transmission matrix}
\label{sec:algebr-T}
To represent the operator ${M}^{\dagger}DM$ as a rotation we first introduce an auxiliary operator
\begin{equation}
  \sigma_z(\psi)\equiv 
  M(\psi)^\dagger\sigma_z{M}(\psi)
\end{equation}
and differentiate it with respect to $\psi$
\begin{align}
  \sigma_z'(\psi)
  &=M^{\dagger}(\psi)\frac{i}{2}[\sigma_x,\sigma_z]M(\psi)
  =\sigma_y(\psi),\\
  \sigma_z''(\psi)
  &={M}^{\dagger}(\psi)\frac{i}{2}[-\sigma_x,\sigma_y]M(\psi)
  =-\sigma_z(\psi),
\end{align}
where we have used commutation properties of Pauli matrices.
The last equation is an ordinary differential equation with the solution
\begin{equation}
  \sigma_z(\psi)=\sigma_z\cos{\psi}+\sigma_y\sin{\psi}.
\end{equation}
Using the unitarity of the matrix $M$, that is ${M}^{\dagger}M=M{M}^{\dagger}=I$, we can write
\begin{equation}
  {M}^{\dagger}\sigma_z^nM=\underbrace{{M}^{\dagger}\sigma_zM\ldots {M}^{\dagger}\sigma_zM}_{n\text{ times}}=\sigma(\psi)^n.
\end{equation}
Presenting an arbitrary function $f(\sigma_z)$ as a Taylor series expansion, it is straightforward to show that 
\begin{equation}
  {M}^{\dagger}f(\sigma_z)M=f({M}^{\dagger}\sigma_zM),
\end{equation}
therefore, we obtain Eq.~(14) for ${M}^{\dagger}\exp(-i\sigma_z\phi/2)M$.

\section{Numerical solution to the inverse design problem}
\label{sec:inverse-design}
To solve the inverse design problem, namely, to find the optimal twisted waveguide approximation to an ideal operator we have devised a numerical optimization problem.
For each ideal rotation operator $T^\mathrm{ideal}$ we must find a twisted waveguide gate $T^\mathrm{twist}$ that maximizes the gate fidelity.
The ideal gate is parametrized with the three Euler angles $\{\vartheta,\varphi,\chi\}$ while the twisted waveguide gate is parametrized with the two parameters: the twist angle $\theta$ and the dimensionless twist length $l=L/L_B$ with $L$ being the (dimensional) twist length and $L_B$ being the linear beat length.
The search of the twisted waveguide parameters was performed in finite parameter space bounded by the maximum twist angle $\theta_\mathrm{max}$ and length $l_\mathrm{max}$.
The performed study can be formally shown using pseudocode
\begin{align}
  \text{For each }&\{\vartheta,\varphi,\chi\}:\nonumber\\
  \text{find }&\underset{\theta,l}{\mathrm{argmax}}[f(\vartheta,\varphi,\chi,\theta,l)],\;|\theta|\leqslant\theta_\mathrm{max},\,l\leqslant l_\mathrm{max}\nonumber\\
  \text{with }&f\left(\vartheta,\varphi,\chi;\theta,l\right)
  =F(T^\mathrm{ideal}(\vartheta,\varphi,\chi),T^\mathrm{twist}(\theta,l))\nonumber\\
\end{align}
The dimensionless length $l$ is related to the angle $\phi$ in~(17) as $\phi=\sqrt{\pi^2l^2+\theta^2}$, the gate fidelity $F=\frac{1}{2}+\frac{1}{12}\sum_{j=x,y,z}\Tr(T\sigma_j{T}^{\dagger}U\sigma_j{U}^{\dagger})$.
The function $f$ is oscillatory with respect to both $\theta$ and $l$ having many local extrema.
Hence to find the global maximum we used a global optimization routine: Differential Evolution Method~\cite{ref:storn1997} available as part of SciPy python library~\cite{ref:scipy2020}.
To ensure convergence to the true global maximum we kept the population size at least as high as $512$ doubling it to $1024$ where needed.
We have performed a series of identical searches but with different design constraints $\theta_\mathrm{max}$ and $l_\mathrm{max}$ and recorded the optimized fidelities for each $\vartheta,\varphi,\chi,\theta_\mathrm{max},l_\mathrm{max}$.
The results of this study are summarized in Figure 3.

\bibliography{references}

\end{document}